# Constraining the origin of giant exoplanets via elemental abundance measurements

H. Knierim, S. Shibata, and R. Helled

Institute for Computational Science, University of Zurich, Winterthurerstr. 90, CH-8057 Zurich, Switzerland, e-mail: henrik.knierim@uzh.ch

**ABSTRACT**

The origin of close-in giant planets is a key open question in planet formation theory. The two leading models are (i) formation at the outer disk followed by migration and (ii) in situ formation. In this work we determine the atmospheric composition of warm Jupiters for both formation scenarios. We perform N-body simulations of planetesimal accretion interior and exterior to the water ice-line for various planetary formation locations, planetary masses, and planetesimal sizes to estimate the accreted heavy-element mass and final planetary composition. We find that the two models differ significantly: migrating giant planets have 2-14 times higher metallicities than planets that form in situ. The ratio between refractories and volatiles is found to be above one for migrating planets but below 0.4 for planets that form in situ. We also identify very different trends between heavy-element enrichment and planetary mass for these two formation mechanisms. While the metallicity of migrating planets is found to increase with decreasing planetary mass, it is about constant for in situ formation. Our study highlights the importance of measuring the atmospheric composition of warm Jupiters and its connection to their formation and evolutionary paths.

**Key words.** Planets and satellites: composition – Planets and satellites: atmospheres – Planets and satellites: formation – Planet-disk interactions – Planets and satellites: fundamental parameters – Planets and satellites: physical evolution

## 1. Introduction

The discovery of 51 Pegasi B, the first exoplanet around a Sun-like star (Mayor & Queloz 1995), came as a genuine surprise: a giant planet on a short-period orbit seemed to be a rare peculiarity. This class of exoplanets, however, soon became widely studied, with well over a thousand confirmed close-in giant exoplanets. Typically, this planetary type is divided into two classes: giant planets with an orbital period below ten days – "hot Jupiters" – and planets with orbital periods longer than 10 days – "warm Jupiters" (WJs).

The origin of giant exoplanets on short-period orbits remains unknown. While various formation pathways have been proposed (see Dawson & Johnson 2018, and references therein), the two main models are (i) formation in the outer disk followed by disk migration (e.g., Lin et al. 1996; Wu & Murray 2003; Beaugé & Nesvorný 2012; Kley & Nelson 2012; Bitsch et al. 2015) and (ii) in situ formation (e.g., Bodenheimer et al. 2000; Ikoma et al. 2001; Boley et al. 2016; Batygin et al. 2016). In the model where the planets form farther out and migrate inward to their current locations (hereafter "model-migration"), a heavy-element[1] core of a few $M_\oplus$ forms in the outer region of the disk and migrates inward. As the core migrates, it accretes large amounts of gas and reaches runaway gas accretion (Pollack et al. 1996). The origin of the planetary embryo is a highly debated subject in itself, where some studies favor planetesimals (e.g., Ida & Lin 2004) and others pebbles (e.g., Bitsch et al. 2015) as the dominant source for core formation. In the in situ formation model (hereafter "model-in situ"), a planetary embryo in the inner region of the disk goes into runaway gas accretion without migrating substantially. The planetary core could come from the outer disk or form in situ.

The expected composition of close-in giant planets differs depending on their formation history and atmospheric chemistry (e.g., Öberg et al. 2011; Madhusudhan 2012; Madhusudhan et al. 2017; Lothringer et al. 2021; Schneider & Bitsch 2021a,b; Hands & Helled 2021; Turrini et al. 2021). With the advent of accurate abundance measurements from the James Webb Space Telescope (JWST) and Ariel missions and ground-based facilities, it will soon be possible to investigate trends in WJ composition. Therefore, predicting observable differences in the enrichment of chemical species is of key importance for constraining planet formation models (e.g., Lothringer et al. 2021; Helled et al. 2022).

In this Letter we determine the atmospheric abundances (focusing on the refractory-to-volatile ratio) of WJs for the two main formation scenarios. We focus on WJs since they are not subject to the radius inflation observed in hot Jupiters, which prevents an accurate estimate of their bulk metallicities (e.g., Bodenheimer et al. 2003; Laughlin et al. 2011). Our Letter is organized as follows: In Sect. 2 we outline the planet formation model. Section 3 presents the results of the enrichment trends. A summary and discussion are presented in Sect. 4.

## 2. Model

### 2.1. Estimating heavy-element enrichment

We used a disk model similar to the one of Hands & Helled (2021); namely, the temperature disk profile is $T(r) = T(1\,\mathrm{AU})/\sqrt{r}$, where $r$ is the distance to the host star and $T(1\,\mathrm{AU}) = 268\,\mathrm{K}$. The solid surface density is given by (Shi-

[1] We refer to elements heavier than helium as heavy elements.

bata et al. 2020)

$$\Sigma_{\rm solid}(r) = Z(r)\Sigma_{\rm gas}(r), \tag{1}$$

where $\Sigma_{\rm gas}$ is the gas surface density according to Lynden-Bell & Pringle (1974) (see Appendix A), and $Z(r)$ is the dust-to-gas mass ratio. Following Hands & Helled (2021), the condensation temperatures are taken from Wood et al. (2019)[2]. For simplicity, we assumed a static disk where the locations of the ice lines remain unchanged throughout the simulation. In reality, their locations change due to the disk's evolution, but typically on timescales longer than the migration timescale considered here. We assumed that the solid material is predominately contained in planetesimals. In that case, $\Sigma_{\rm solid}$ is equivalent to the surface density of planetesimals (for details, see Sect. 2.2). The planetesimals' composition follows the host star's composition of the solid species at their respective location. For example, a planetesimal with a semimajor axis $a$ has an oxygen abundance of $X_{\star,\rm O}/Z(a)$ if $a$ is exterior to the water ice-line and zero if $a$ is interior to the water ice-line. The $X_{\star,\rm O}$ is the stellar oxygen abundance, which we assumed to be proto-solar (Lodders 2020). While Hands & Helled (2021) focused on the bulk captured mass at two locations, interior and exterior to the water ice-line, we account for the composition of each individual planetesimal (see Appendix C).

Next, we considered the two formation scenarios: model-migration and model-in situ. In both cases, we assumed that the giant planet is fully formed and has the same composition as the gas at the planet's initial location. Consequently, we did not consider the effects of gas enrichment due to pebble evaporation (e.g., Schneider & Bitsch 2021a,b). Instead, we assumed that any additional enrichment is solely a result of late-stage planetesimal accretion.

The refractory enrichment, $E_{\rm ref}$ (and analogously the volatile enrichment, $E_{\rm vol}$), used here is an expansion of the model by Hands & Helled (2021). Namely, the enrichment is given by the ratio

$$E_{\rm ref} = \frac{M_{\rm p,ref}}{M_{\star,\rm ref}}, \tag{2}$$

where $M_{\rm p,ref}$ is the mass of refractory elements in the planet (including contributions from gas accretion, planetesimal accretion, and potentially a core) and $M_{\star,\rm ref}$ is the mass of refractories assuming stellar abundance, which is set to be proto-solar (Lodders 2020). In this study we evaluated potassium and oxygen enrichments as proxies for refractories and volatiles, respectively.

### 2.2. Formation in the outer disk followed by migration

In model-migration, we performed N-body simulations of planetesimal accretion by a migrating protoplanet (type II) around a solar-mass star. We determined the accreted planetesimal mass analogous to Shibata et al. (2020) and terminated the simulation once the planet reached the disk's inner edge. As the planet migrated, planetesimal accretion was controlled by mean motion resonances that can lead to significant accretion in the region referred to as the sweet spot for planetesimal accretion (SSP; Shibata et al. 2022). The location of the SSP is controlled by the disk gas drag timescale, $\tau_{\rm aero,0}$, and the migration timescale, $\tau_{\rm tide,a}$: a shorter (longer) migration timescale would give the same result as smaller (larger) planetesimals as long as $\tau_{\rm aero,0}/\tau_{\rm tide,a}$ remains constant (Shibata et al. 2022).

[2] See Fig. 1 in Hands & Helled (2021).

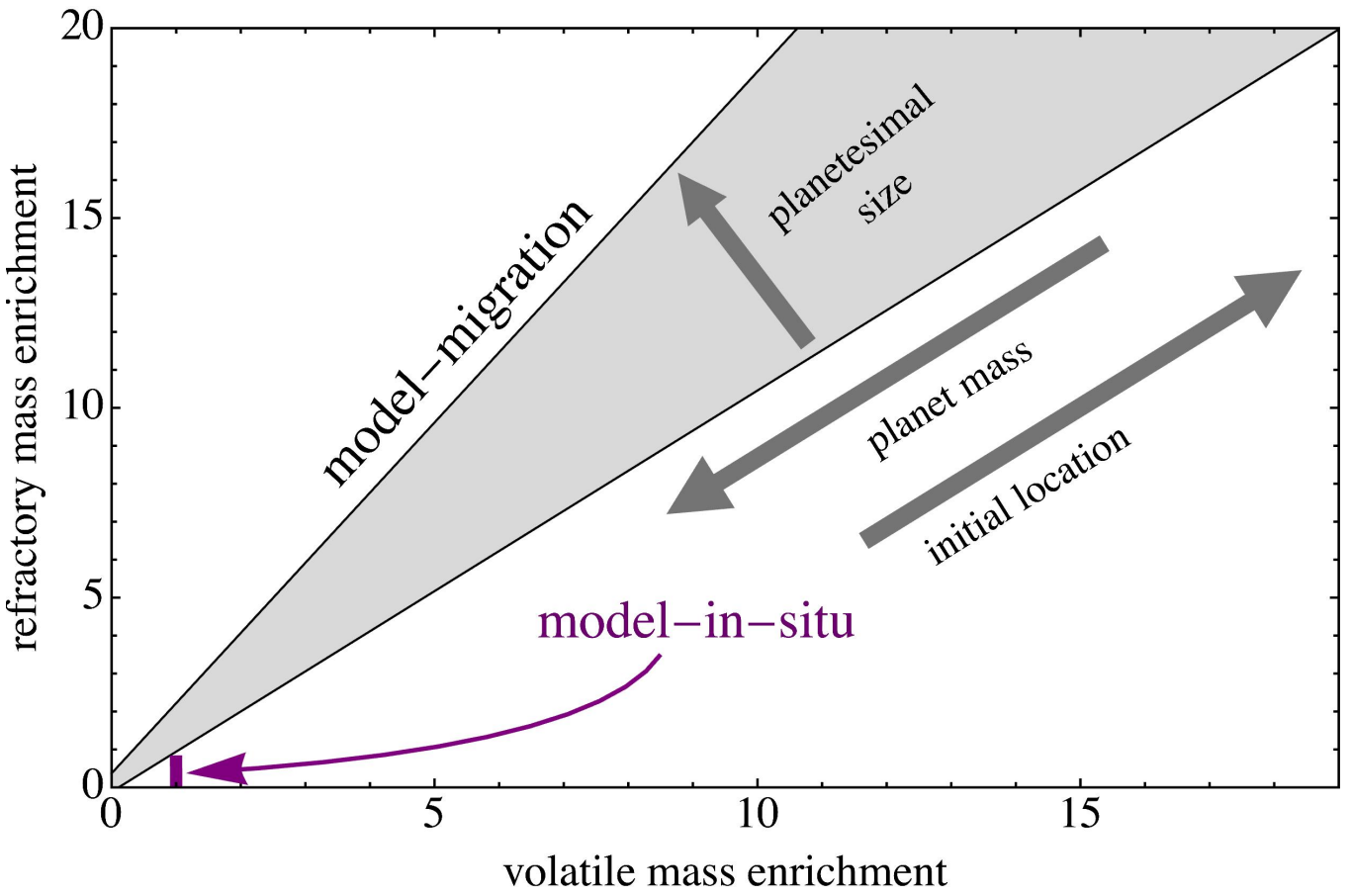

**Fig. 1.** Refractory mass enrichment vs. volatile mass enrichment (Eq. 2) for model-in situ (purple) and model-migration (gray). The gray region indicates the predicted parameter range and the gray arrows the trend with planetesimal size, planetary mass, and formation location.

Consequently, to estimate the possible enrichment for model-migration, we assumed a fixed migration timescale at $\tau_{\rm tide,a} = 10^5$ yr and three sizes for the planetesimals, $R_{\rm pl}$: $10^5$, $10^6$, and $10^7$ cm. In addition, we investigated three different initial formation locations, $a_{\rm p,0}$: 10, 20, and 30 AU. More details on the numerical setup can be found in Appendix B, in Table D.1, and in Shibata et al. (2020, 2022).

### 2.3. In situ formation

In model-in situ, the WJ is fixed at 0.5 AU and captures planetesimals in its feeding zone $[r_{\rm min}, r_{\rm max}]$, set by the Hill radius, $r_{\rm Hill}$, according to $r_{\rm min/max} = r_{\rm p} \mp 2\sqrt{3}r_{\rm Hill}$, where $r_{\rm p}$ is the planet's semimajor axis. The captured mass, $M_{\rm cap}$, can be written as

$$M_{\rm cap} = k \cdot 2\pi \int_{r_{\rm min}}^{r_{\rm max}} r\Sigma_{\rm solid}(r)\,{\rm d}r, \tag{3}$$

where $k \approx 0.3$ (Shibata & Ikoma 2019). If $k$ is higher, as suggested by other studies (Podolak et al. 2020, for example, find $k$ values between 0.2 and 0.55), the captured mass increases accordingly. Close to the host star the feeding zone is typically small, and Eq. 3 can be approximated as

$$M_{\rm cap} \sim 2\pi k r_{\rm p}\Sigma_{\rm solid}(r_{\rm max} - r_{\rm min}). \tag{4}$$

Since $r_{\rm Hill} \propto M_{\rm p}^{1/3}$, $M_{\rm cap} \propto M_{\rm p}^{1/3}$.

## 3. Results

### 3.1. Absolute element enrichment

Figure 1 compares the inferred enrichment of refractories and volatiles of model-migration and model-in situ (for details, see Tables D.1 and D.2). We find that in model-migration the enrichment in both refractory and volatile elements can be increased by over an order of magnitude above stellar composition. Moreover, volatile and refractory materials are found to be up to ∼ 20 and ∼ 70 times more enriched compared to model-in situ, respectively. This is expected since the mass reservoir for planets that form in situ is significantly smaller than that of migrating planets.

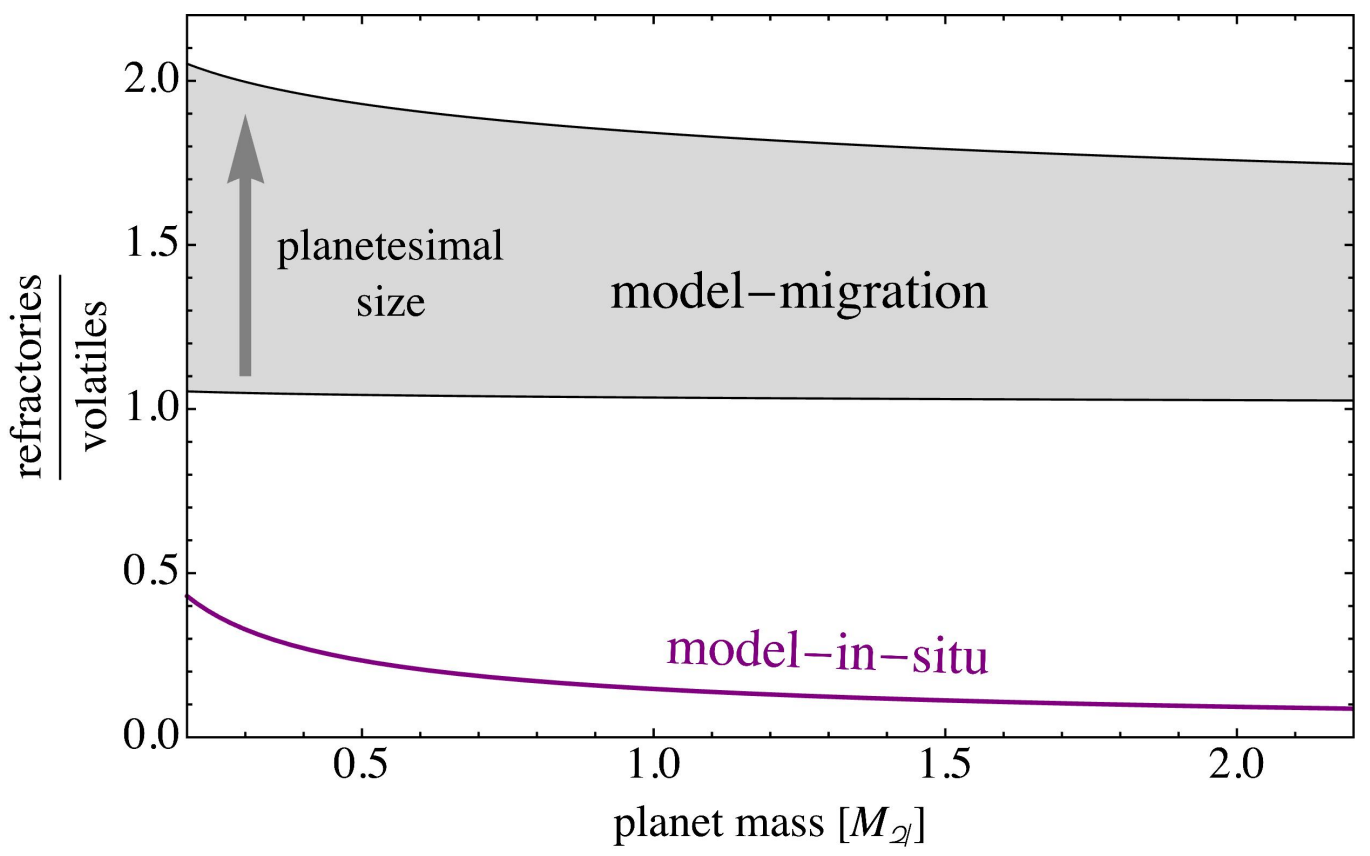

**Fig. 2.** Normalized refractory-to-volatile ratio vs. planetary mass. The gray region shows the parameter space we predict for model-migration, and the purple line shows model-in-situ. The gray arrow indicates the expected scaling with planetesimal size.

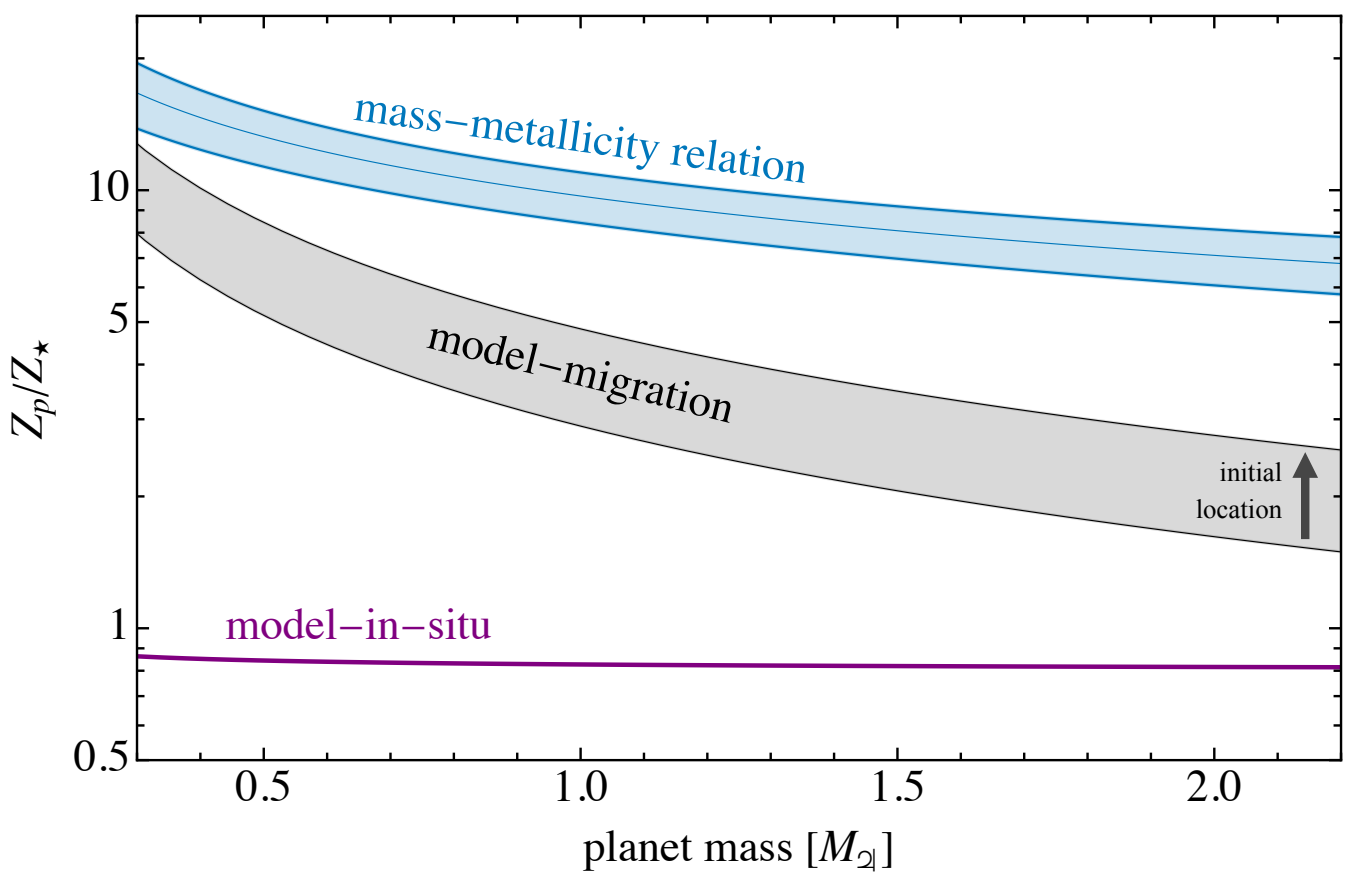

**Fig. 3.** Mass-metallicity relation for model-in situ (purple), model-migration (gray), and the result from Thorngren et al. (2016) (blue).

Giant planets that form in situ can only capture planetesimals in their feeding zone, which is not being replenished. Although the captured mass increases with planetary mass (Eq. 4), the overall enrichment in refractory material decreases because $E_{\rm ref} \propto M_{\rm cap}/M_{\rm p} \propto M_{\rm p}^{-2/3}$. On the other hand, migrating giant planets can capture planetesimals in resonance. Once the migrating planet enters the SSP, the resonances can be broken and the planet can capture the accumulated planetesimals. The SSP of lower-mass planets is closer to the star, and therefore less massive planets can accrete more planetesimals (Shibata et al. 2022). Thus, less massive planets are more strongly enriched. Additionally, larger planetesimals experience less gas drag. Since gas drag is crucial for breaking the resonances, their SSP moves inward (i.e., interior to the water ice-line), increasing the refractory-to-volatile ratio in the planets. We can therefore conclude that close-in giant planets that are enriched in heavy elements formed in the outer disk and migrated inward, while close-in giant planets that have low bulk metallicities could have formed in situ.

### 3.2. Refractory-to-volatile ratio

Figure 2 shows the inferred refractory-to-volatile enrichment for both formation scenarios. It should be noted that here we refer to the ratio of the mass enrichment (Eq. 2), and not their respective mass. In model-migration the refractory-to-volatile ratio is between 1 and 2, unlike in model-in-situ, where the refractory-to-volatile ratio is below 0.4. The low refractory-to-volatile ratio in model-in situ is expected since giant planets that form in situ can accrete only a small number of (refractory-rich) planetesimals but have massive gaseous envelopes with stellar volatile abundance. In both models the refractory-to-volatile ratio decreases with increasing planetary mass.

Planets that form in the outer disk and migrate begin with volatile-poor envelopes, but can accrete a significant mass of planetesimals consisting of a mixture of volatiles and refractories. As discussed above, larger planetesimals tend to be accreted interior to the water ice-line, increasing the refractory-to-volatile ratio. This assumes, however, that planetesimals outside the water ice-line are volatile-rich. The accreted planetesimals could be on highly eccentric orbits due to mean motion resonances and therefore experience strong ablation from the ambient disk gas, which can lead to the evaporation of volatile materials (e.g., Pollack et al. 1986; Podolak et al. 1988; Tanaka et al. 2013; Eriksson et al. 2021). As a result, the refractory-to-volatile ratio shown in Fig. 2 for model-migration represents the lower bound. Overall, our results clearly predict that migrating giant planets are more enriched in refractories than volatiles, while WJs that form in situ are expected to be volatile-rich.

### 3.3. Mass-metallicity relation

Figure 3 shows the inferred planetary metallicity versus planetary mass for both models. The planetary bulk metallicity in model-migration is found to be significantly higher than in model-in situ. We find that for model-migration, WJs are enriched with heavy elements by up to a factor of $\sim$ 10 compared to their host star for planets below $0.5\,M_{♃}$ and $\sim 2-5$ for planets beyond $\sim 1\,M_{♃}$. In contrast, in model-in situ WJs have $Z_{\rm p}/Z_{\rm star} \sim 0.8$ for the entire mass range we consider.

In general, we predict that WJs that form in situ have substellar metallicity, whereas migrating WJs are enriched with heavy elements. Given that many of the WJs are expected to be enriched in heavy elements (e.g., Thorngren et al. 2016; Müller et al. 2020), formation at the outer disk followed by migration is the favorable explanation (Shibata et al. 2020, 2022).

The metallicity of giant planets in model-migration increases with increasing initial formation location simply because more planetesimals can be captured in resonance during migration (Shibata et al. 2020, 2022). In contrast, the dominating source for enrichment in model-in situ is the volatile-rich accreted gas since planetesimal accretion is negligible. As a result, model-in situ cannot reproduce the mass-metallicity relation of giant exoplanets (e.g., Thorngren et al. 2016), while model-migration reaches sufficiently high metallicities. Since the inferred slope of model-migration depends on the migration timescale, which is unknown, and on the existing large theoretical uncertainties in the derived mass-metallicity relation (Müller et al. 2020), this discrepancy is not significant. It should also be noted that we do not include the heavy-element core in our simulations. A massive heavy-element core could contribute substantially to the bulk metallicity. However, although both mass-metallicity curves might shift upward, the overall different trends are expected to persist.

### 3.4. Connection to observations

Retrieving the atmospheric abundances of exoplanets is a primary endeavor of exoplanet science. Currently, characterization is still limited to the hottest and largest class of exoplanets: hot Jupiters (e.g., Welbanks et al. 2019; Yip et al. 2021; Nikolov et al. 2022; McGruder et al. 2022). Although we focus on WJs, measurements of hot Jupiters can still point toward compositional trends. In Fig. 4 we compare atmospheric abundance measurements of hot Jupiters with predictions from model-in situ and model-migration. Unfortunately, the uncertainties of the retrieved atmospheric abundances are still too large to discriminate between the two models.

For example, different atmospheric retrievals of WASP-96b have large uncertainties that are consistent with both formation scenarios (right side of Fig. 4). Moreover, while potassium abundances are more consistent with model-in situ, sodium abundances favor model-migration, demonstrating that a more complete approach is needed to understand those measurements in detail (see Sect. 4).

We also note that neither formation scenario can reproduce the extremely low water abundance of HD209458b, $E_{\mathrm{H_2O}} = 0.08^{+0.09}_{-0.04}$, or HD189733b, $E_{\mathrm{H_2O}} = 0.031^{+0.04}_{-0.017}$ (converted to mass enrichment from Welbanks et al. 2019). Furthermore, the water abundance presented in Fig. 4 could be different from the oxygen abundance (and volatile content) within the planet. This is because there could be other sources of oxygen, and the ratio between water and oxygen abundance depends on various complex physical and chemical processes and their interplay. Nonetheless, Hands & Helled (2021) suggest that a planet forming close to the water ice-line (5 AU in their case) and migrating rapidly ($10^3$ yr) can have substellar water abundances while retaining super-stellar refractory abundances. It is yet to be explored whether such rapid migration is feasible. For longer migration timescales, Hands & Helled (2021) arrive at results consistent with this study. This highlights the importance of understanding planet migration for constraining planet formation and predicting planetary composition. Rather than those somewhat limited case-by-case studies, however, larger samples with more accurate atmospheric abundances are clearly needed. Future measurements from JWST and Ariel could therefore help link exoplanet composition with formation in a statistically significant manner.

## 4. Summary and discussion

We have investigated the difference in the predicted bulk and atmospheric composition of giant planets assuming formation in the outer disk followed by migration and in situ formation. We compared WJs between 0.26 and 2.1 $M_{♃}$ growing in situ at 0.5 AU with their migrating counterparts. For migrating WJs, we considered various formation locations (10, 20, and 30 AU) and the accretion of planetesimals of different sizes (1 – 100 km). Using a simple chemical disk model, we determined the refractory and volatile content in the planets, as well as their metallicity. We find that there is a clear and robust difference between the two formation scenarios in the predicted enrichment trend. Moreover, we find that this dichotomy persists over a wide range of planetary masses. Our key conclusions can be summarized as follows:

1. Migrating WJs have super-solar bulk metallicities, while WJs that form in situ have subsolar to solar bulk metallicities
2. Giant planets that form in the outer disk and migrate inward are predicted to be more enriched, by a factor of ten or more, than giant planets that form in situ.
3. The inferred normalized refractory-to-volatile ratio for model-migration is between 1 and 2, and below 0.4 for model-in situ.
4. The bulk metallicity of giant planets that form in the outer disk and migrate inward is $\sim 5 - 15$ times larger in comparison to giant planets that form in situ for low-mass WJs and $\sim 2 - 6$ times higher for high-mass WJs.

While our conclusions seem to be robust and are rather insensitive to the model assumptions, our model is simple and does not account for all the possible physical processes.

First, we neglected mixing or settling of the accreted material and assumed that the accreted material remains in the outer atmosphere. In addition, we did not consider physical or chemical processes that could affect the atmospheric abundance, such as cloud formation, gravitational settling, and atmospheric loss. If such post-formation mechanisms are dominant, the observed abundances could differ significantly from the bulk composition. This would alter the observations of both the mass enrichment and the refractory-to-volatile ratio.

Second, our disk chemistry model is very simplistic. Here we focused on the water ice-line and the transition from volatile-rich to volatile-poor solids; however, it is clear that including more tracer species and considering more complex disk chemistry is desirable. In particular, the composition of WJs in model-in situ is governed by the composition of the accreted gas. In this study we assume that the gas interior to the water ice-line has a stellar volatile abundance, but in fact the gas composition could range between substellar and super-stellar. For example, pebble drift followed by evaporation could enrich the inner disk with volatile materials (e.g., Schneider & Bitsch 2021b; Aguichine et al. 2022). Conversely, gas supplied by the outer disk could be volatile-poor due to the formation of solids that are accreted by growing objects in the outer disk. Therefore, it is clear that further investigations focusing on the gas composition for in situ formation is desirable, and we hope to address it in future research. Nevertheless, the overall dichotomy between model-in situ and model-migration is expected to remain.

Third, the planetesimal's size was fixed throughout the simulations. Planetesimal collisions and ablation may change the planetesimals' size and composition. This in turn can affect the dynamics and, hence, the accretion efficiency of planetesimals.

Finally, the formation model itself does not include certain physical processes, such as gas accretion during planetary migration, pebble accretion, or giant impacts. Schneider & Bitsch (2021a,b) considered planet formation in the pebble accretion framework. While their model can also produce the high bulk metallicities of WJs, they arrive at considerably greater volatile abundances, leading to refractory-to-volatile ratios below one. Thus, this ratio could also be used to discriminate between planetesimal-dominated and pebble-dominated formation models, as well as the physical and chemical processes that are considered. It is therefore clear that further investigations and more sophisticated models are needed.

Despite the importance of a more comprehensive analysis, the overall dichotomy between formation in the outer disk followed by migration and in situ formation is expected to remain. We therefore emphasize the great potential of upcoming atmospheric measurements of giant exoplanets for a range of planetary masses, as these could reveal key information on planetary origin. We look forward to having results from JWST, *Ariel*, and ground-based observations.

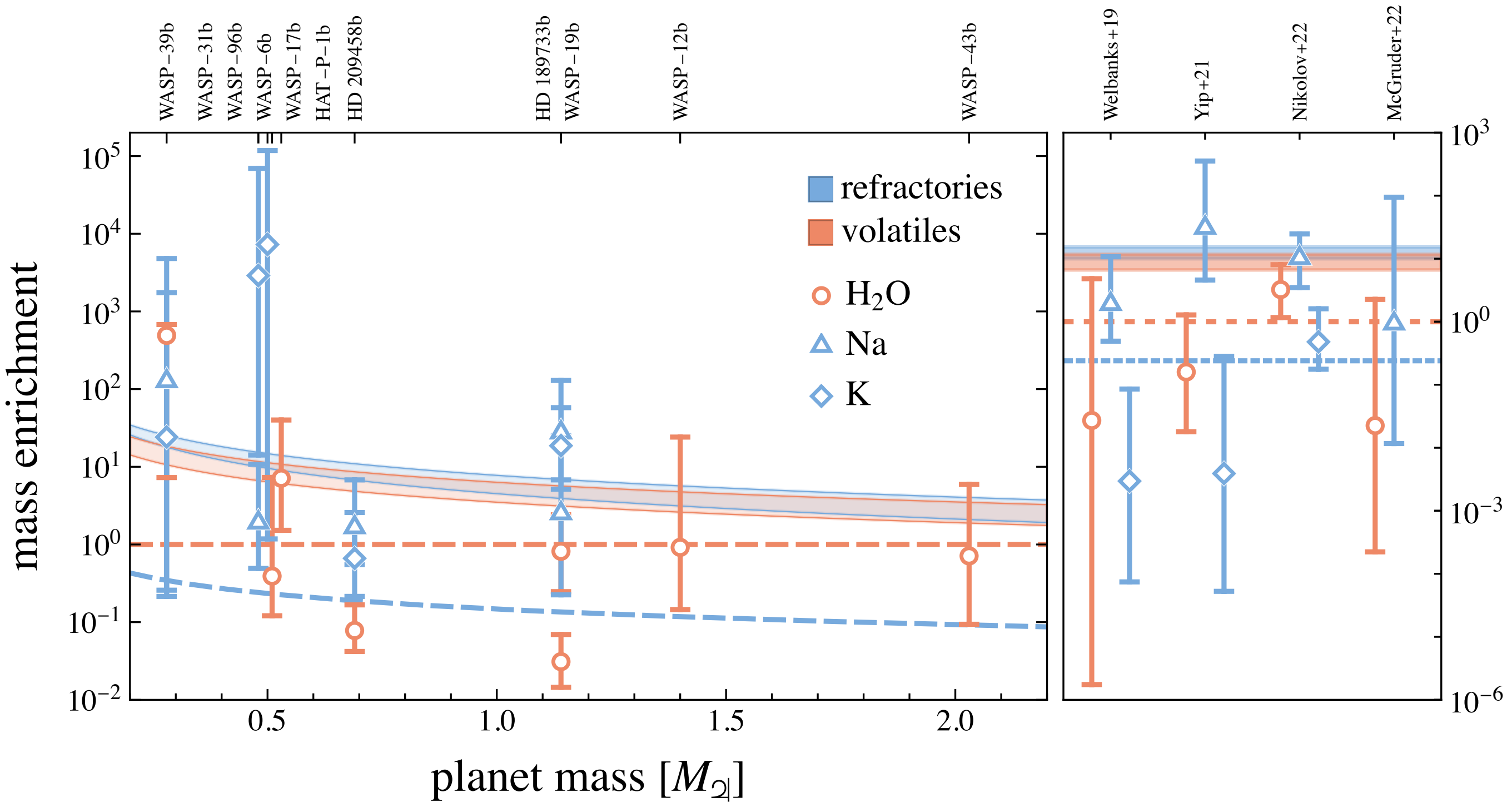

**Fig. 4.** Comparison of atmospheric retrieval studies with this work. *Left:* Refractory enrichment (blue) and volatile enrichment (orange) for model-migration (colored regions) and model-in situ (dashed lines). The orange circles, blue triangles, and blue diamonds show the retrieved mass abundances of Welbanks et al. (2019) for $H_2O$, Na, and K, respectively. We note that WASP-31b and WASP-96b have the same mass of 0.48 $M_♃$. For this mass, the potassium measurement corresponds to WASP-31b and the sodium measurement corresponds to WASP-96b. *Right:* Same as left but for four retrievals of WASP-96b: Welbanks et al. (2019), Yip et al. (2021), Nikolov et al. (2022), and McGruder et al. (2022). We note that the orange circle for Nikolov et al. (2022) shows the oxygen abundance and not the water abundance. Also, we have converted the volume mixing ratios from the atmospheric retrieval studies to mass enrichment.

*Acknowledgements.* We thank an anonymous referee for valuable comments. We also acknowledge support from the Swiss National Science Foundation (SNSF) under grant `200020_188460`. Some of this work has been carried out within the framework of the National Centre of Competence in Research PlanetS supported by the Swiss National Science Foundation under grants 51NF40_182901 and 51NF40_205606. Numerical computations were partly carried out on the Cray XC50 at the Center for Computational Astrophysics, National Astronomical Observatory of Japan.

## Appendix A: Self-similar disk profile

The self-similar solution for the gas surface density is given by

$$\Sigma_{\rm gas} = \frac{M_{\rm disk}}{2\pi R_m^2}\left(\frac{r}{R_m}\right)^{-1}\tilde{t}^{-\frac{3}{2}}\exp\left(-\frac{r}{\tilde{t}R_m}\right), \tag{A.1}$$

where $R_m = 50\,\mathrm{AU}$ is the radial scale length, $M_{\rm disk} = 0.1M_\odot$ is the initial disk's mass, and

$$\tilde{t} = \frac{t_0}{\tau_s} + 1. \tag{A.2}$$

Here, $t_0 = 3\times10^6$ yr is the starting time of our simulation and

$$\tau_s = \frac{R_m^2}{3\nu(R_m)}, \tag{A.3}$$

where we use the $\alpha$-disk approximation (Shakura & Sunyaev 1973) for $\alpha = 10^{-2}$ to compute the viscosity, $\nu$.

## Appendix B: Numerical setup for model-migration

In protoplanetary disks, the orbits of planetesimals are damped by the disk's gas drag. We adopted the gas drag model of Adachi et al. (1976), where the typical timescale of disk gas drag, $\tau_{\rm aero,0}$, is given by

$$\tau_{\rm aero,0} = \frac{8R_{\rm pl}\rho_{\rm solid}}{3C_{\rm d}v_{\rm K}\rho_{\rm gas}}, \tag{B.1}$$

where $C_{\rm d}$ is the nondimensional drag coefficient, $\rho_{\rm solid}$ is the planetesimal's density, $\rho_{\rm gas}$ is the gas density, and $v_{\rm K}$ is the Keplerian velocity. The $\rho_{\rm solid}$ is set to 2.4 and $1.0\,\mathrm{g\,cm^{-3}}$ for planetesimals interior and exterior to the water ice-line, respectively. The values of $C_{\rm d}$ and $\rho_{\rm gas}$ are calculated from the disk model.

We terminated the N-body simulations once the protoplanet reached the disk inner edge, $a_{\rm p,f}$. Planetesimal accretion rapidly decreases after the protoplanet enters the region interior to the SSP. Thus, $a_{\rm p,f}$ does not affect the numerical results as long as $a_{\rm p,f}$ is smaller than the inner edge of the SSP. In our simulations, we set $a_{\rm p,f} = 0.2\,\mathrm{AU}$ for $R_{\rm pl} = 10^7\,\mathrm{cm}$ and $a_{\rm p,f} = 0.5\,\mathrm{AU}$ for $R_{\rm pl} = 10^5\,\mathrm{cm}$ and $R_{\rm pl} = 10^6\,\mathrm{cm}$. The planetesimals were set as test particles that were distributed between $a_{\rm pl,in}$ and $a_{\rm pl,out}$. We used $N_{\rm p}$ test particles in each simulation. The values of $N_{\rm p}$, $a_{\rm pl,in}$, and $a_{\rm pl,out}$ depend on $a_{\rm p,0}$ (see Table D.1). Further details on the model can be found in Shibata et al. (2020).

## Appendix C: Enrichment functions

The enrichment, $E$, of a planet for a specific element is given by

$$E_{\rm el} = \frac{M_{\rm cap,el}}{X_{\odot,\rm el}(M_{\rm p}+M_{\rm cap})} + \frac{X_{\rm gas,el}}{X_{\odot,\rm el}}\frac{M_{\rm p}}{M_{\rm p}+M_{\rm cap}}, \tag{C.1}$$

where $M_{\rm cap,el}$ is the captured mass of the element, $X_{\odot,\rm el}$ is the proto-solar mass abundance of the element, $M_{\rm p}$ is the planet's mass before accreting planetesimals, $M_{\rm cap}$ is the total captured mass, and $X_{\rm gas,el}$ is the mass abundance of the element in the gaseous disk. The captured mass of a specific element is the sum of the captured planetesimals, $m_i$, weighted by the mass abundance at the capture location, $r_i$:

$$M_{\rm cap,el} = \sum_i \frac{X_{\rm el}(r_i)}{Z(r_i)}m_i. \tag{C.2}$$

Likewise, the total captured mass is simply $M_{\rm cap} = \sum_i m_i$.

We applied a simple composition model where once a planetesimal crosses an ice-line of a chemical element or molecule, we assumed that the respective element or molecule completely evaporates:

$$X_{\rm el}(r_i) = \Theta(T_{\rm cond,el} - T(r_i))X_{\odot,\rm el}, \tag{C.3}$$

where $\Theta(T)$ is the Heaviside function. Similarly, the local metallicity, $Z(r_i)$, is the sum of the icy material:

$$Z(r_i) = \sum_{\rm el} X_{\rm el}(r_i). \tag{C.4}$$

Consequently, Eq. C.1 simplifies to

$$E_{\rm el} = \frac{1}{M_{\rm p}+M_{\rm cap}}\left(\sum_i \frac{\Theta[T_{\rm cond,el} - T(r_i)]m_i}{Z(r_i)} + E_{\rm gas,el}M_{\rm p}\right), \tag{C.5}$$

where

$$E_{\rm gas,el} = \Theta[T_{\rm cond,el} - T(a_{\rm p,0})] \tag{C.6}$$

is the gas enrichment.

## Appendix D: Detailed simulation results

All the results we obtain from the N-body simulations are summarized in Table D.1. The parameter ranges for model-migration we present in the main text are fits to these data. For an easy comparison, results from model-in situ are shown in Table D.2.

**Table D.1.** Simulation results of model-migration.

| $M_p[M_{♃}]$ | $R_{pl}$[cm] | $a_{p,0}$[AU] | $a_{p,f}$[AU] | $N_p$ | $a_{pl,in}$[AU] | $a_{pl,out}$[AU] | $m_{cap}[M_{♃}]$ | $E_{ref}$ | $E_{vol}$ | $E_{ref}/E_{vol}$ | $Z_p/Z_{star}$ |
|---|---|---|---|---|---|---|---|---|---|---|---|
| 0.26 | $10^6$ | 20 | 0.5 | 19 200 | 0.3 | 26. | 0.05 | 18.40 | 13.81 | 1.33 | 10.2 |
| 0.52 | $10^6$ | 20 | 0.5 | 19 200 | 0.3 | 26. | 0.08 | 14.71 | 10.60 | 1.39 | 8.1 |
| 1.05 | $10^6$ | 20 | 0.5 | 19 200 | 0.3 | 26. | 0.07 | 6.09 | 5.70 | 1.07 | 4.1 |
| 2.1 | $10^6$ | 20 | 0.5 | 19 200 | 0.3 | 26. | 0.06 | 2.86 | 2.39 | 1.20 | 2.0 |
| 0.26 | $10^7$ | 20 | 0.2 | 19 200 | 0.1 | 26. | 0.05 | 19.35 | 10.36 | 1.87 | 8.8 |
| 0.52 | $10^7$ | 20 | 0.2 | 19 200 | 0.1 | 26. | 0.06 | 13.24 | 7.14 | 1.85 | 6.2 |
| 1.05 | $10^7$ | 20 | 0.2 | 19 200 | 0.1 | 26. | 0.07 | 9.47 | 4.03 | 2.35 | 4.0 |
| 2.1 | $10^7$ | 20 | 0.2 | 19 200 | 0.1 | 26. | 0.06 | 3.29 | 2.20 | 1.49 | 2.0 |
| 0.26 | $10^5$ | 20 | 0.5 | 19 200 | 0.3 | 26. | 0.07 | 19.06 | 18.09 | 1.05 | 12.5 |
| 0.52 | $10^5$ | 20 | 0.5 | 19 200 | 0.3 | 26. | 0.07 | 10.10 | 9.77 | 1.03 | 6.8 |
| 1.05 | $10^5$ | 20 | 0.5 | 19 200 | 0.3 | 26. | 0.06 | 4.42 | 4.28 | 1.03 | 3.2 |
| 2.1 | $10^5$ | 20 | 0.5 | 19 200 | 0.3 | 26. | 0.06 | 2.33 | 2.26 | 1.03 | 1.8 |
| 0.26 | $10^6$ | 10 | 0.5 | 9 600 | 0.3 | 13. | 0.04 | 15.99 | 10.24 | 1.56 | 8.1 |
| 0.52 | $10^6$ | 10 | 0.5 | 9 600 | 0.3 | 13. | 0.04 | 9.12 | 5.29 | 1.72 | 4.5 |
| 1.05 | $10^6$ | 10 | 0.5 | 9 600 | 0.3 | 13. | 0.04 | 3.76 | 3.48 | 1.08 | 2.7 |
| 2.1 | $10^6$ | 10 | 0.5 | 9 600 | 0.3 | 13. | 0.05 | 2.03 | 1.76 | 1.15 | 1.6 |
| 0.26 | $10^6$ | 30 | 0.5 | 28 800 | 0.3 | 39. | 0.06 | 19.48 | 14.9 | 1.31 | 11.0 |
| 0.52 | $10^6$ | 30 | 0.5 | 28 800 | 0.3 | 39. | 0.08 | 14.65 | 10.05 | 1.46 | 7.9 |
| 1.05 | $10^6$ | 30 | 0.5 | 28 800 | 0.3 | 39. | 0.09 | 7.32 | 6.91 | 1.06 | 4.9 |
| 2.1 | $10^6$ | 30 | 0.5 | 28 800 | 0.3 | 39. | 0.08 | 3.43 | 2.85 | 1.20 | 2.3 |

**Notes.** The planetary masses correspond to $0.25 \times 10^{-3}, 0.5 \times 10^{-3}, 1 \times 10^{-3}$, and $2 \times 10^{-3}\, M_{\odot}$, respectively.

**Table D.2.** Simulation results of model-in situ.

| $M_p[M_{♃}]$ | $m_{cap}[M_{\oplus}]$ | $E_{ref}$ | $E_{vol}$ | $Z_p/Z_{star}$ |
|---|---|---|---|---|
| 0.26 | 0.11 | 0.36 | 1.0 | 0.87 |
| 0.52 | 0.13 | 0.23 | 1.0 | 0.84 |
| 1.05 | 0.17 | 0.14 | 1.0 | 0.83 |
| 2.1 | 0.21 | 0.09 | 1.0 | 0.82 |

**Notes.** The planetary masses correspond to $0.25 \times 10^{-3}, 0.5 \times 10^{-3}, 1 \times 10^{-3}$, and $2 \times 10^{-3}\, M_{\odot}$, respectively.